\documentclass[aps,prl,superscriptaddress,preprint]{revtex4-2}

\usepackage{graphicx}% Include figure files
\usepackage{dcolumn}% Align table columns on decimal point
\usepackage{bm}% bold math
\usepackage{mathtools}

\begin{document}

\preprint{APS/123-QED}
\title{Influence of Kerr Anisotropy in Parametric Amplification}
%\title{Kerr instability amplification with polarization rotation}

\author{Nathan G. Drouillard}
\affiliation{Dept. of Physics, University of Windsor, Windsor ON N9B 3P4 Canada}
\author{Fadi Farook}
\affiliation{Dept. of Physics, University of Toronto, Toronto ON M5S 1A7 Canada}%
\author{Meerna Albert}%
\affiliation{Dept. of Physics, University of Windsor, Windsor ON N9B 3P4 Canada}
\author{Rachel Durling}%
\affiliation{Dept. of Physics, University of Windsor, Windsor ON N9B 3P4 Canada}
\author{Jordan Saad}%
\affiliation{Dept. of Physics, University of Windsor, Windsor ON N9B 3P4 Canada}
\author{Jeffrey G. Rau}
\affiliation{Dept. of Physics, University of Windsor, Windsor ON N9B 3P4 Canada}
\author{TJ Hammond}
\affiliation{Dept. of Physics, University of Windsor, Windsor ON N9B 3P4 Canada}
 %\affiliation[Also at ]{Physics Department, XYZ University.}%Lines break automatically or can be forced with \\
 \email{tj.hammond@uwindsor.ca}
%\affiliation[1]{%

\date{\today}% It is always \today, today,

\begin{abstract}
Four-wave parametric amplification can be extended to the TW/cm$^2$ regime using femtosecond pump pulses to amplify nearly octave spanning pulses with gain $> 20$~mm$^{-1}$, which we call Kerr instability amplification. Cross-polarized wave generation exploits Kerr anisotropy to induce a transient intensity-dependent polarization evolution. In this work, we combine Kerr instability amplification with cross-polarized wave generation to simultaneously amplify and rotate the output polarization of a signal beam, and we explore laser and crystal parameters to control the resulting polarization. In 1~mm MgO(100), we amplify linearly polarized light by $2000\times$ orthogonal to the pump and seed polarization. The parametric amplification and polarization rotation offers excellent pulse contrast enhancement for future high-power laser systems. Furthermore, the polarization provides an additional observable to study the nonlinear dynamics occuring in this extreme ultrafast light-matter interaction. 
\end{abstract}

\maketitle

%\tableofcontents

\section{Introduction}
%Optical parametric amplifiers, based on the second-order nonlinear susceptibility, $\chi^{(2)}$, have driven many advances in ultrafast physics \cite{ManzoniJOpt2016}. In particular, their large tunability makes them ideal for femtosecond spectroscopy \cite{XXX}, and high contrast pulses are ideal for studying strong-field effects such as high harmonic generation spectroscopy and attosecond (1~as = 10$^{-18}$~s) science \cite{YYY, ZZZ}.  High harmonic generation in condensed matter has led to a search for crystals that can withstand extreme intensities, such as magnesium oxide (MgO), which can be pumped by femtosecond (1~fs = $10^{-15}$~s) pulses above 10~TW/cm$^2$ without optical damage \cite{KorobenkoOE2019, HeinrichNatComm2021}. 

Because of its moderate Kerr nonlinearity \cite{AdairPRB1989} and dispersion \cite{StephensJRNBS1952}, lack of first-order Raman response \cite{MansonPRB1971}, and high damage threshold MgO is an ideal material for studying ultrafast strong-field effects in condensed matter such as high harmonic generation and attosecond science \cite{KorobenkoOE2019, YouNatPhys2017, HeinrichNatComm2021, ZhangOQE2021, HussainPRA2022}. MgO has recently extended four-wave mixing to this high-intensity regime, called Kerr instability amplification (KIA), leading to large amplification of nearly octave-spanning pulses \cite{Ghosh2024super}. This broad spectrum amplification and tunability has applications in nonlinear optics and femtosecond (1 fs = 10$^{-15}$ s) stimulated Raman spectroscopy, where the broad amplified bandwidth leads to a probe spectrum spanning the entire Raman spectrum of interest in a single measurement \cite{DrouillardAS2025}.

Four-wave mixing in optical fibers leads to modulational polarization instabilities, a phenomenon that has been extensively studied \cite{WabnitzPRA1988}. These instabilities can be influenced by various factors, including parametric amplification and Raman scattering \cite{MillotJOSAB2014}. Both the signal amplification and polarization characteristics can depend on the pump configuration \cite{FreitasJOSAB2007}, and controlled using orthogonally polarized pump waves \cite{LinJOSAB2004}. In highly birefringent fibers, the fiber’s intrinsic birefringence plays a dominant role in determining the polarization behavior \cite{GuasoniJOSAB2012}.

Although MgO is not birefringent, its cubic crystal structure exhibits polarization-dependence in the higher-order $\chi^{(3)}$ nonlinearity. This nonlinear anisotropy can lead to cross-polarized wave generation (XPW), an intensity-dependent orthogonal polarization rotation \cite{MinkovskiOL2002, MinkovskiJOSAB2004, JullienJOSAB2005}. XPW can be used to improve pulse contrast and pulse and spectral shaping \cite{JullienOL2005, JullienAPB2007, ZhaoAIP2022}, useful in high-intensity fs laser systems \cite{RamirezJOSAB2013, BuberlOE2016}. Furthermore, we expect that this nonlinear polarization rotation will affect the efficiency and resulting polarization in high harmonic generation at these extreme intensities \cite{YouNatPhys2017, AllegreOL2025}.

We find that when the intense femtosecond pump pulses are polarized along the MgO(100) direction, we observe significant polarization sensitivity in the amplified beam, but not when polarized in the MgO(110) direction. To explain this nonlinear polarization rotation, we extend four-wave mixing in the extreme intensity regime to include polarization dependence.

\section{Theory}

Due to the nonlinear polarization rotation, we include the tensorial nature of the Kerr nonlinearity. The nonlinear material polarization, $\vec P^{NL}$, depends on the field, $\vec E$, and including only the third-order nonlinearity is
\begin{align}
    P_i^{NL}(\vec E) = \epsilon_0 \sum_{jkl} \chi^{(3)}_{ijkl} E_j E_k E_l
\end{align}
where $i,j,k,l$ are the $x,y,z$ Cartesian coordinates. Because the relative pump-seed angle is $3^\circ$ (within the sample), we assume that the nonlinearity is only in the $xy$ plane to simplify the calculations \cite{KourtevJOSAB2009}. The crystal structure of MgO is rock-salt cubic with space group \textit{m3m}, as shown in Fig. \ref{fig:setupcrystal}(a). In this space group, the only non-zero tensor components of the third-order susceptibility are $\chi^{(3)}_{xxyy}$, $\chi^{(3)}_{xyyx}$, $\chi^{(3)}_{xyxy}$, and $\chi^{(3)}_{xxxx}$ \cite{ButcherandCotter}. Including both transverse polarizations \cite{AgrawalNFO, Jullienthesis}, the nonlinear polarization is then
\begin{subequations}
\begin{align}
    P_x^{NL} = \frac{3}{8} \epsilon_0 \Big[ \chi^{(3)}_{xxxx} |E_x|^2 E_x + 2 \chi^{(3)}_{xyyx} |E_y|^2 E_x + \chi^{(3)}_{xxyy} E_y^2 E_x^* \Big] \\
    P_y^{NL} = \frac{3}{8} \epsilon_0 \Big[ \chi^{(3)}_{xxxx} |E_y|^2 E_y + 2 \chi^{(3)}_{xyyx} |E_x|^2 E_y + \chi^{(3)}_{xxyy} E_x^2 E_y^* \Big] 
\end{align}
\end{subequations}
The first term corresponds to self-phase modulation, while the second term leads to cross-polarization coupling, and the third term is degenerate four-wave mixing. 

We assume that the nonlinear susceptibility is frequency independent such that $\chi^{(3)}_{xyyx} = \chi^{(3)}_{xyxy}$ \cite{HutchingsJOSAB1997}. When far from one and two photon resonances, we can further simplify using $\chi^{(3)}_{xyyx} = \chi^{(3)}_{xxyy}$, and thus we only need the two values for the nonlinear susceptibility, $\chi^{(3)}_{xxyy}$ and $\chi^{(3)}_{xxxx}$. This $\chi^{(3)}_{xxyy}$ leads to cross-phase modulation and energy exchange between polarizations \cite{JullienJOSAB2005}. 

To simulate the pulse propagation, we use Forward Maxwell's Equation \cite{HusakouPRL2001}:
\begin{align}
    \frac{\partial \vec E(\vec r, \omega)}{\partial z} = i \frac{\omega}{c} \Big[ n(\omega) - n_g \Big] \vec E(\vec r, \omega) + i \frac{c}{2 n(\omega) \omega} \nabla_\perp^2 \vec E(\vec r, \omega) + i \frac{\mu_0 \omega c}{2 n(\omega)} \vec P(\vec r, \omega)
\end{align}
which allows us to include the frequency-dependent refractive index $n(\omega)$ over the broad bandwidth of interest. The group index, $n_g$ is determined at the pump wavelength; $c$ is the speed of light and $\mu_0$ is the vacuum permeability. We simulate only one transverse dimension, that is $\nabla_\perp^2 = \partial^2/\partial x^2$. Further details on the simulation are discussed in \cite{Ghosh2024super, Ghosh2024single, DrouillardPRA2024}. 

For our simulations, we match the experimental conditions discussed below, but increase the pump beam waist from 85 (experimental) to $200$~\textmu m and (full width at half maximum) pulse duration from $\Delta t_{FWHM} = 110$ (experimental) to 200~fs to improve spatial separation and decrease the walk-off effects. We note that the nonlinear Kerr coefficient for MgO(100) has been previously determined to be $n_2(100) = 3.90 \times 10^{-20}$~m$^2$/W \cite{AdairPRB1989}. Using the relation that 
\begin{align}
n_2 = \frac{3}{4} \frac{\chi^{(3)}}{\epsilon_0 c n^2}
\end{align}
where $n \approx 1.73$ \cite{StephensJRNBS1952} and $c$ is the speed of light in vacuum, then the third order nonlinear susceptibility is  $\chi^{(3)}_{xxxx} = 4.12 \times 10^{-22}$~m$^2$/V$^2$, and $\chi^{(3)}_{xxyy} = 0.482 \chi^{(3)}_{xxxx}$. 

We have previously approximated Kerr instability amplification in the plane-wave monochromatic case as $I(L) = I_0 \cosh^2 \bigl( g \frac{L}{2} \bigr)$ \cite{Ghosh2024single}, where
\begin{align}
g \approx \frac{2\pi}{\lambda_p} \frac{n_2}{n_p} I_p \Bigl[2 - \cosh\Bigl(\frac{\lambda_p}{\lambda_s} - 1\Bigr) \Bigr] \label{eq:gain}
\end{align}
is the gain, $I_0$ is the initial seed intensity, $\lambda_p$ and $\lambda_s$ are the pump and seed wavelengths, respectively, $n_2$ is the nonlinear Kerr coefficient, $n_p = n(\lambda_p)$ is the index of refraction at the pump wavelength, $I_p$ is the peak pump intensity, and $L$ is the crystal length. Because the Kerr coefficient depends on the crystal orientation \cite{AdairPRB1989},
\begin{align}
\chi^{(3)}_{100} &= \chi^{(3)}_{xxxx} \\
\chi^{(3)}_{110} &= \frac{\chi^{(3)}_{xxxx} +3\chi^{(3)}_{xxyy}}{2} \label{eq:tensor}
\end{align}
we can use the orientation-dependent gain to retrieve changes in the Kerr coefficient.

Phase matching for KIA becomes intensity dependent due to the intense pump pulse. The internal optimum phase matching angle becomes
\begin{equation}
\cos\theta_s = \frac{4 (n_p^2 - n_2 I_p) \omega_p^2 + n_s^2 \omega_s^2 - n_i^2 \omega_i^2}{4 n_s \omega_s \omega_p \sqrt{n_p^2 - n_2 I_p}}. \label{eq:phasecurve}
\end{equation}
where subscripts $p,s,i$ are the pump, signal, and idler values, respectively. We can relate this internal angle to the experimental angle through Snell's law. The relative pump-seed angle of $5.3^\circ$ optimizes phase matching at 15~TW/cm$^2$, and we do not expect amplification until $>9$~TW/cm$^2$ at this angle \cite{NesrallahOptica2018, Ghosh2024super}. 

\section{Methods}

\subsection{Experimental Setup}

\begin{figure}[htb]
\includegraphics[width=1\columnwidth]{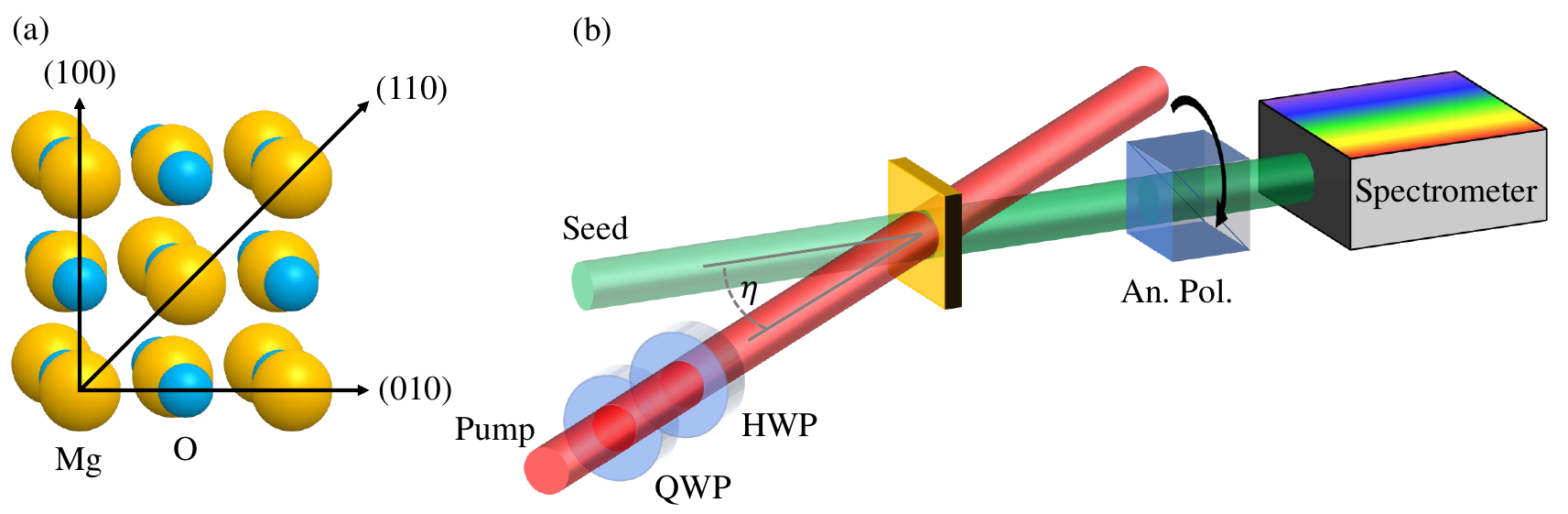}
\caption{(a) The rock-salt MgO crystal structure with \textit{m3m} symmetry. The beam propagates in the (001) direction; vertical polarization is in the (100) direction. Rotating the crystal by 45 degrees allows us to access the (110) orientation. (b) The setup for measuring the polarization rotation in Kerr instability amplification with the MgO crystal in the (100) orientation. We control the relative incident pump-seed angle ($\eta$) and the polarization of the pump independently. The rotating analyzing polarizer (An. Pol.) and spectrometer measure the polarization. }\label{fig:setupcrystal}
\centering
\end{figure}

The setup for the KIA polarization measurements is shown in Fig. \ref{fig:setupcrystal}(b). We generate the supercontinuum seed in 3~mm thick sapphire, spanning from below $500$~nm to greater than $1000$~nm; we use a visible filter to remove the 785~nm laser peak (for further details, see \cite{Ghosh2024super}), a neutral density (ND) wheel to control the power (to avoid saturation effects), and lenses to collimate and refocus the beam. The optical elements between the supercontinuum generation and the MgO crystal introduce significant dispersion (approximately 1600~fs$^2$) to the seed \cite{DrouillardPRA2024}. Because of the supercontinuum chirp, temporal overlap of the seed is limited to $\sim50$~nm bandwidth with the 110~fs pump pulse. We then combine the seed with the intense pump (relative external angle $\eta=5.3 \pm 0.2^{\circ}$) in a MgO(100) cut crystal (crystal thicknesses 0.2, 0.5, and 1.0~mm). Keeping the pump polarization vertical, we can rotate the MgO crystal to orient along the (100) to the (110) direction. After amplification, the pump is spatially separated and we pass the amplified beam through a wire grid polarizer (on a computer-controlled rotation mount) and focus it into a visible spectrometer (OceanOptics Flame-S). At zero degrees, the wire grid polarizer blocks the vertically polarized light. We adjust the pump polarization with the $\lambda/4$ (QWP) and $\lambda/2$ (HWP) wave plates; in this configuration we ensure linearly polarized pump light while reversing the wave plate order allows for measuring ellipticity dependence.

\begin{figure}[htb]
\includegraphics[width=1\columnwidth]{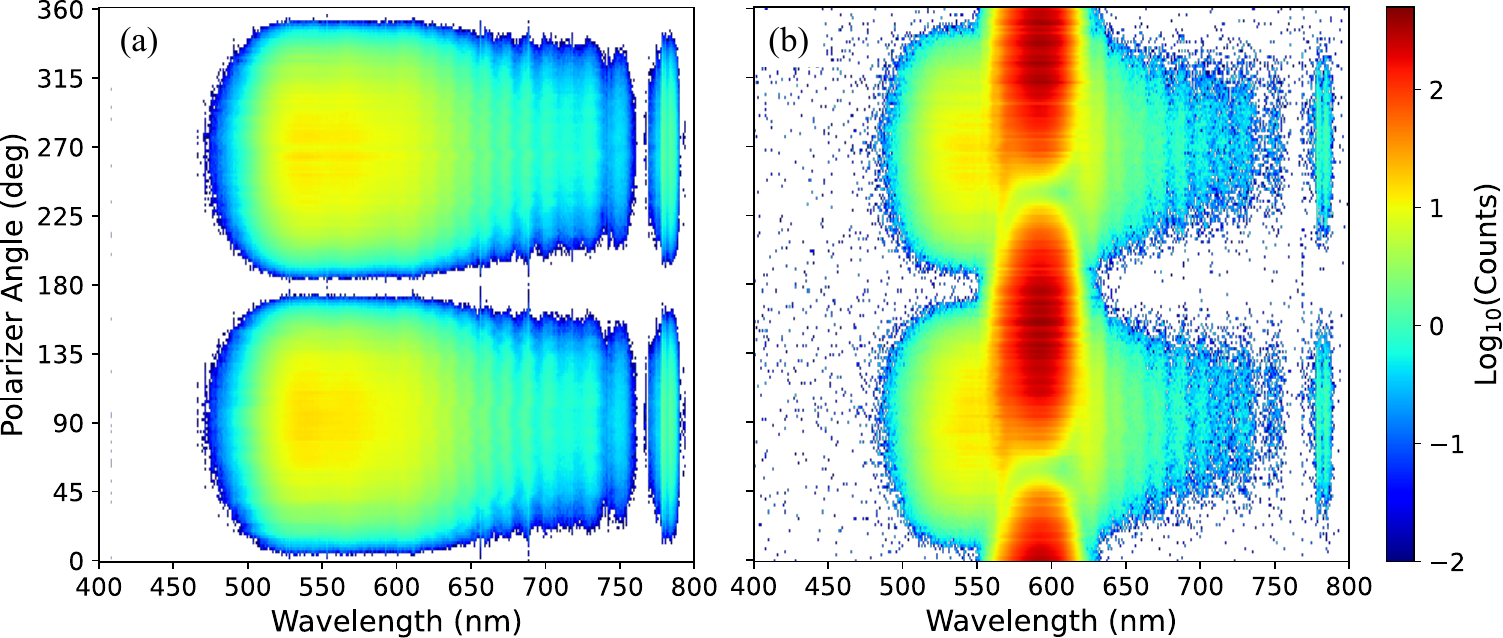}
\caption{(a) Linearly polarized supercontinuum seed spectrum filtered to span the visible region. The seed is linearly polarized in the vertical (90$^\circ$) direction. (b) Measured spectrum with amplification around 595~nm showing polarization rotation. The two figures share common axes.}\label{fig:spectra}
\centering
\end{figure}

We calibrate the amplification by changing the integration time and the transmission through the ND wheel. We record the polarized seed spectrum at maximum ND wheel transmission with sufficient integration time to minimize noise. Then, with the pump amplifying the seed, we decrease the ND wheel to minimum transmission and adjust the integration time to maximize the spectrometer dynamic range while avoiding saturation. We then record the seed spectrum (without pump) again at this integration time, using the ratio of the maximum counts to calibrate the ND wheel transmission; an example seed spectrum is shown in Fig. \ref{fig:spectra}(a), demonstrating that the seed spectrum is linearly vertically polarized. We calculate the amplification factor as the maximum counts for the amplified spectrum compared to the seed at that wavelength.

To measure the angle, a nonlinear curve fitting algorithm \cite{curve_fit} fits our polarization-dependent spectra to $I(\theta) = A \cos^2 ( \theta + \phi_0) + B$, where $\theta$ is the polarizer angle, the fitting parameters are $A$ and $B$ for the amplitudes, and $\phi_0$ is for the phase, which we denote as the measured angle of the amplified beam.

\subsection{Amplification and Polarization Rotation}

With the pump and seed polarization vertical and when the crystal is in the (100) orientation, we observe a rotation in the amplified spectrum, as shown in Fig. \ref{fig:spectra}(b). In this case, we selected the spectral region around 595~nm to be amplified by tuning the pump-seed temporal overlap. While the spectrum is amplified by a factor of 20, it is also significantly rotated by $60^{\circ}$. The peak pump intensity is estimated to be $I_p = 7\times10^{16}$~m$^2$/W in a 0.5~mm thick crystal. This significant rotation even with weak amplification demonstrates the sensitivity of the resulting polarization to the pump.

%The amplified spectrum bandwidth is limited by the dispersion of the supercontinuum seed optics, which significantly chirps the seed (estimated dispersion 1600~fs$^2$ \cite{DrouillardPRA2024}); this chirp allows us to select the amplified wavelength by adjusting the relative delay of the pump and seed. 

\begin{figure}[htb]
\includegraphics[width=0.95\columnwidth]{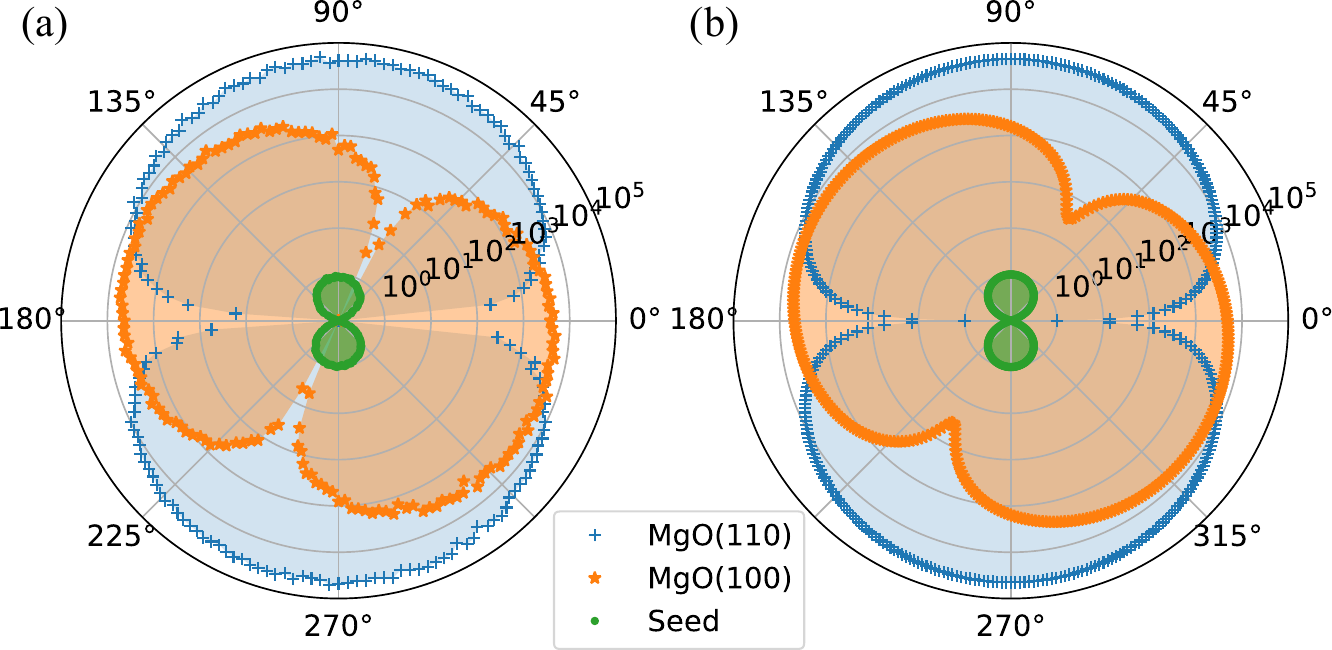}
\caption{Polar plot (radius is amplification magnitude order) of crystal orientation dependence on polarization rotation and amplification. The seed (green circle) is linearly polarized with maximum transmission normalized to unity. (a) At high pump intensity, the amplification in MgO(100) (orange star) shows the polarization rotation ($5500\times$ amplification), while MgO(110) (blue plus) ($45 000\times$ amplification) does not. (b) Simulation of amplification and rotation with crystal orientation dependence.}\label{fig:polaramplification}
\centering
\end{figure}

We now show the effect of the crystal axis orientation with large amplification in Fig. \ref{fig:polaramplification}(a). The seed lineout (green) is minimum at 0$^{\circ}$ (the uncertainty in our relative alignment of the polarizations, along with the polarizer and crystal orientation, is approximately $4^{\circ}$), indicating linearly vertically polarized light. We normalize the maximum seed spectrum to unity, and we measure the amplification as relative to the seed spectrum at the wavelength of maximum gain. In the strong pump regime, in the MgO(100) crystal orientation (orange), we clearly observe a polarization rotation. The polarization remains linear with the axis ratio $> 1000:1$, but rotated by 65$^{\circ}$ relative to the seed. The maximum amplification of 5500$\times$ at 600~nm corresponds to a gain of $g(100) = 20.0$~mm$^{-1}$ for the 0.5~mm thick MgO crystal. 

We then rotate the MgO crystal by 45$^{\circ}$ so that the polarization is along the (110) crystal axis. In this case, the polarization of the amplified beam (blue) matches that of the seed with the minimum near zero. In this case, the maximum amplification is $45 000\times$ at 675~nm, corresponding to a gain of $g(110) = 24.2$~mm$^{-1}$. We find that this polarization is linear and aligns to the (110) crystal axis even under small perturbations \cite{HutchingsJOSAB1997}.

In this intensity regime, we measure the amplified pulse energy to be $340 \pm 30$~nJ in the (100) configuration, and $3.2 \pm 0.3$~\textmu J  in the (110) configuration. With the pump power at 250~\textmu J, the estimated conversion efficiency is 1.3\%, or typical of the first stage of an OPA \cite{ManzoniJOpt2016}. We find that the amplification is typically an order of magnitude more for the (110) than the (100) orientation. 

Since changing the orientation changes only the nonlinear Kerr coefficient, $g(110)/g(100)  = n_2(110)/n_2(100) = 1.21$. We find better agreement in our simulations with experiment when $n_2(100) = 3.0 \times 10^{-20}$~m$^2$/W; using the above gain ratio leads to $\chi^{(3)}_{xxyy} = 0.54 \chi^{(3)}_{xxxx}$ (i.e. $n_2(110) = 3.7 \times 10^{-20}$~m$^2$/W). Using these values of the angle-dependent Kerr nonlinearity, we show good agreement with simulations in Fig. \ref{fig:polaramplification}(b). We note that although the amplification factor agrees well with the experimentally determined values ($6300\times$ for MgO(100), $45 000\times$ for MgO(110)), the initial pump polarization must be rotated to 14$^\circ$ in the MgO(100) case to rotate the polarization by $61^\circ$, significantly more than we experimentally measured (pump polarization sensitivity is discussed below) and the amplified field is not as linearly polarized (axis ratio $\sim 200:1$). Although the gain at such high intensities will require a more complete physical understanding, including multiphoton absorption anisotropy, higher order nonlinear susceptibilities, and plasma effects, we use these Kerr coefficients to inform our simulations. Measuring the Kerr coefficient anisotropy at these extreme intensities requires further study \cite{DrouillardKerrmeasurements}.

\section{Pulse Contrast}

Improving pulse contrast of fs laser sources through nonlinear polarization rotation has been well explored in the single beam case \cite{JullienAPB2006, RicciRMP2013}. High-power sources demand high pulse contrast to pre- and post-pulses to drive the nonlinear process of interest, such as preventing ionization from the pre-pulse that degrades phase matching in high harmonic generation \cite{FattahiOE2016}. The single-beam setup benefits from its simplicity, but is limited to approximately 10\% efficiency (given as the amount of power transmitted through orthogonal polarizers). Adding more crystals \cite{JullienOE2006} and changing the crystal cut \cite{CanovaAPL2008} can increase the efficiency to nearly 30\%. Although KIA requires two beams, and is thus more complicated than single-beam cross-polarized wave generation, parametric amplification further improves the pulse contrast.

\begin{figure}[htb]
\includegraphics[width=1.0\columnwidth]{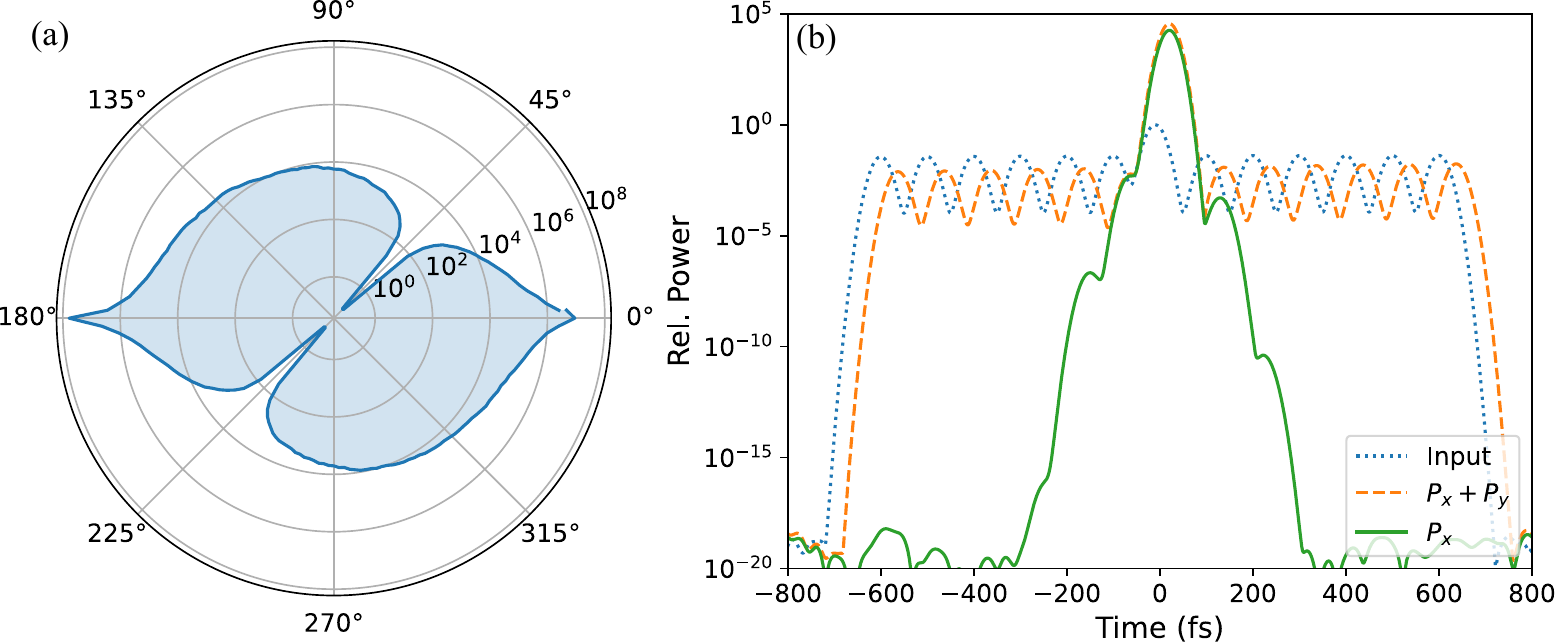}
\caption{The nonlinear polarization rotation improves pulse contrast. (a) We measure the contrast as the ratio of the amplified beam to the seed, measuring more than six orders of magnitude improvement. (b) Simulation of the improved pulse contrast. Initial seed (blue dotted) has several pre- and post-pulses, separated by 100 fs; amplified pulse is rotated by 45$^\circ$ (orange dash is the $xy$ direction); the resulting pulse in $x$ polarization (green solid) has four orders of magnitude amplification and 15 orders of magnitude extinction relative to background.}\label{fig:contrast}
\centering
\end{figure}

We note that the intensities used here are more than an order of magnitude higher than in BaF$_2$, a material with higher anisotropy previously used for improving pulse contrast \cite{RamirezJOSAB2013}. Although MgO has a lower anisotropy as defined as $\sigma = 1 - 3 \chi^{(3)}_{xxyy}/\chi^{(3)}_{xxxx}$ \cite{Jullienthesis}, the higher peak intensities available for MgO may lead to more efficient nonlinear rotation. 

Because we see maximum amplification when the pump is polarized along the MgO(110) axis, we use this orientation for measuring the improved pulse contrast. The seed remains vertically polarized while the pump polarization ($I_p = 15$~TW/cm$^2$) is rotated to 45$^\circ$ (along the MgO(110) axis). We then compare the ratio of the amplified beam to the seed at each polarizer angle, as shown in Fig. \ref{fig:contrast}(a). The measured polarization contrast reaches more than six orders of magnitude, limited mainly by the dynamic range of the spectrometer.

We also simulate the pulse contrast in the time domain, as shown in Fig.  \ref{fig:contrast}(b). The input pulse is a series of 20~fs pulses centered at 600~nm, separated by 100~fs, all vertically polarized (blue dotted). The pump polarized at 45$^\circ$ along the (110) axis with peak intensity $I_p = 15$~TW/cm$^2$ and 100~fs pulse duration centered at 800~nm. In this configuration, the pump pulse aligns the amplified field along the (110) axis. The output pulse along the (110) direction (orange dashed) is amplified by 36 000$\times$ at the peak of the pump (delayed by 20~fs due to propagation effects). Isolating the \textit{x} portion of the output beam halves the power, but significantly improves the pulse contrast (green solid). Although the pulse contrast will be experimentally limited by the quality of the polarizers, we expect that the large amplification will improve the contrast by nearly five orders of magnitude compared to the single beam case.

\section{Power and Length Scaling}

%Although the power scaling of KIA has been previously investigated \cite{JACOL2021}, we include it here because both the polarization angle and the amplification are fit in this measurement. 

To perform the power scaling, we rotate a half wave plate with a polarizer in the pump beam while keeping all other parameters fixed. For this measurement, the crystal is 0.2~mm thick MgO in (100) configuration. The thinner crystal avoids self-focusing and propagation effects, allowing for a higher peak intensity before observing damage.

\begin{figure}[htb]
\includegraphics[width=1\columnwidth]{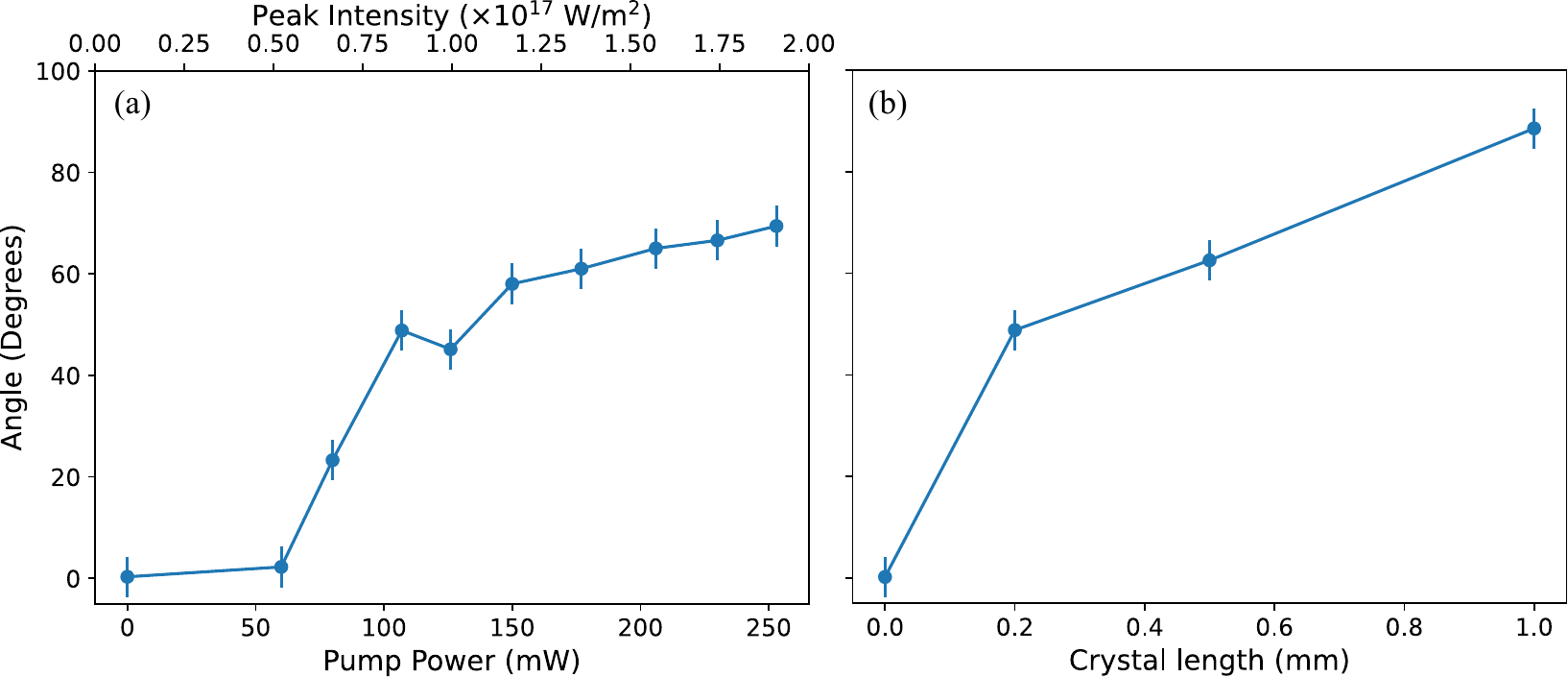}
\caption{(a) Pump power scaling of rotation in 0.2~mm thick MgO(100). The rotation saturates around $70^\circ$. (b) The polarization rotation for 1.0~mm is nearly orthogonal to the seed.}\label{fig:scaling}
\centering
\end{figure}

When the seed is without pump, the amplification is unity and the rotation angle is zero. We could not detect an influence of the pump below 50~mW. We estimate that for a pump power of 127~mW the peak intensity is 10~TW/cm$^2$, which is just above the threshold for amplification according to eq. \ref{eq:phasecurve}. Interestingly, we observe rotation before the onset of amplification. We find that there is an intensity above which we see a sudden change in the polarization, as shown in Fig. \ref{fig:scaling}(a). Although there is an obvious intensity dependence to the threshold, we find from our simulations that the non-collinear phase matching also contributes to the polarization rotation. At higher intensities, this polarization rotation then saturates around $70^{\circ}$. The maximum power we could impinge on the crystal was 250~mW (approximate peak intensity 19~TW/cm$^2$) before damage, above which amplification significantly deteriorated. 

We also measure the effect of the crystal length on the nonlinear rotation. Because of self-focusing effects in the longer crystal, we decrease the pump intensity to 7~TW/cm$^2$ for these measurements. In Fig. \ref{fig:scaling}(b), we plot the measured amplified beam polarization for the different crystal thicknesses. The nonlinear polarization rotation again shows a high sensitivity for the shorter crystal, and then only modestly continues to rotate. We measure $88 \pm 4$ degrees with amplification of $2000\times$ in the 1~mm case, enabling the generation of orthogonally polarized femtosecond pulses relative to the seed. 

We note the similarities in the two scaling cases. The amount of nonlinearity experienced in the material depends on the Kerr coefficient, the intensity, and the crystal length, which we combine into a parameter $\gamma = \frac{2\pi}{\lambda_p} n_2 I_p L$. There is a threshold value of $\gamma$, where the resulting polarization rapidly changes from 0 to above 45$^\circ$, when $\gamma \sim \pi$. Above this value, the rotation saturates and is repeatedly measured to be 60\---70$^\circ$. 

Above 15~TW/cm$^2$ we expect plasma to become a dominant contributor to the nonlinearity \cite{McDonaldPRL2017}, however we do not observe significant change in the amplification character. At these intensities, we expect the conduction band population to be $>10\%$ leading to a decrease in the nonlinear index through the relation \cite{BoydNonlinearOptics}
\begin{align}
n(I_p) \approx n_0 + n_2 I_p - \frac{N e^2}{2 \epsilon_0 m^* \omega_p^2}
\end{align}
where the free electron density $N$ depends on the intensity-dependent ionization, $e$ is the electron charge,  $m^*$ is the effective electron mass, and $\omega_p$ is the pump frequency. Such a high free electron density should significantly decrease the amplification, but is not readily observed. It may be possible to separate the higher-order nonlinearities and plasma effects by comparing length scaling to power scaling since a longer crystal will compensate for lower intensities, and thus avoid plasma effects. However, we found for crystals longer that 0.5~mm, pulse propagation effects such as walkoff and group delay limit pump-seed overlap.

\section{Initial Polarization Dependence}

\begin{figure}[htb]
\includegraphics[width=0.6\columnwidth]{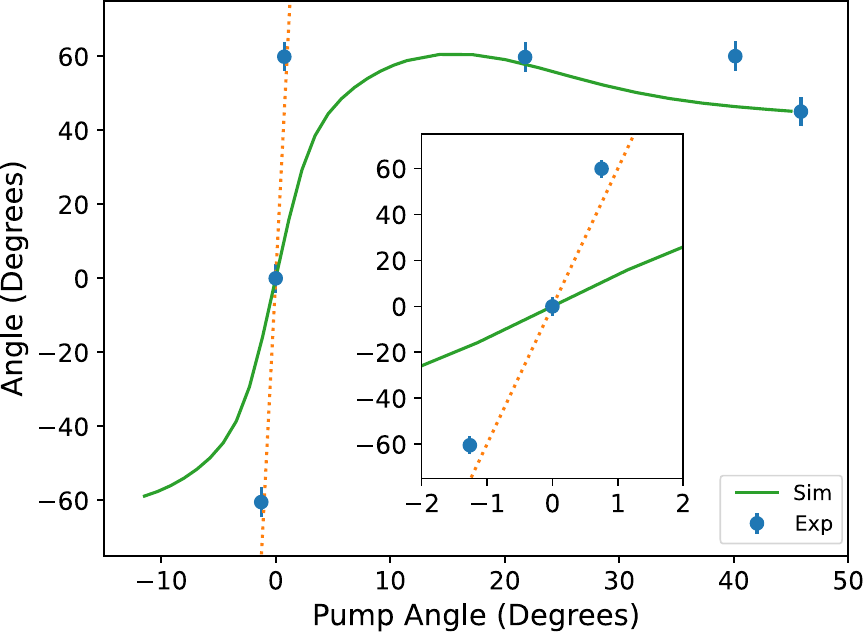}
\caption{Measured amplified beam polarization depends on the initial pump polarization. The polarization is much more sensitive to the pump polarization about 0 degrees in experiment (blue dots) than simulation (green solid); at 45 degrees the amplified beam follows the MgO(110) axis. Inset: zoom in on the polarization measurement about 0 degrees for the pump laser; orange dotted line to guide the eye. }\label{fig:ogpolarization}
\centering
\end{figure}

We also explore the influence of the initial polarization on the amplified beam polarization, as shown in Fig. \ref{fig:ogpolarization}. We find that the resulting polarization is independent of the seed polarization, in agreement with simulations. We also find qualitative agreement between experiment and simulations (in both cases, $I_p = 15$~TW/cm$^2$) for the pump polarization dependence. However, we find experimentally that the resulting measured polarization angle is highly sensitive to the initial pump polarization, such that any change by $\pm 1^\circ$ rotates the polarization by $\pm 60^\circ$ (highlighted by the inset). We find the simulations give a much smoother continuous distribution of resulting polarization rotations. Furthermore, we measure that this $60^\circ$ angle is a constant, independent of the pump polarization from 1 to $40^\circ$, and at $45^\circ$ we again get alignment along the (110) axis. In our simulations, the maximum rotation is at $60^\circ$, but smoothly decreases to $45^\circ$.

We can also measure the polarization linearity of the amplified beam. Again, we find qualitative agreement between simulation and experiment. We measure linearly polarized light when the pump is at 0 and $45^\circ$, and the amplified beam is elliptically polarized from 1 to $22^\circ$, however the degree of ellipticity is less experimentally, as shown in Fig. \ref{fig:polaramplification}.

Higher order nonlinearities and plasma formation may account for discrepancies between our simulations and experiment. Further investigations are required to study the nonlinear rotation of the pump and determine the improvement in pulse contrast in a single-beam case. In fact, our seed follows the polarization of the pump pulse, acting as a probe of the rotation that the pump pulse undergoes while propagating through the material. The polarization sensitivity and gain of KIA may offer opportunities to measure such rich dynamics.

\section{Conclusions}
We have observed significant polarization rotation in Kerr instability amplification when the pump is polarized near the MgO(100) direction. We find that the polarization rotation is pump intensity and crystal length dependent, which allows for control over the polarization rotation. We expect that this nonlinear polarization rotation will benefit high-power systems by significantly improving pulse contrast. Although our simulations show good qualitative agreement with experiment,  the differences such as pump polarization sensitivity, may offer insights into higher order effects contributing to this extreme photonics environment.

\begin{acknowledgments}
We acknowledge funding from Natural Sciences and Engineering Research Council of Canada (RGPIN-2019-06877) and the University of Windsor Xcellerate grant (5218522). TJH thanks Thomas Brabec for useful discussions.
\end{acknowledgments}

%\nocite{*}

\bibliography{KIA_anisotropy}% Produces the bibliography via BibTeX.

\end{document}